\documentclass[aps,prd,nofootinbib,floatfix,twocolumn]{revtex4}

\usepackage{amsmath,amssymb,color,graphicx,bm}
\usepackage{physics}

\usepackage{hyperref}

\definecolor{red}{rgb}{0.8,0,0}
\definecolor{RED}{rgb}{0.8,0,0}
\definecolor{violet}{rgb}{0.4,0,0.4}
\definecolor{green}{rgb}{0,0.5,0.0}
\definecolor{GREEN}{rgb}{0,0.5,0.0}
\definecolor{navy}{rgb}{0.0,0.0,0.6}
\definecolor{orange}{rgb}{0.8,0.2,0.0}
\definecolor{blue}{rgb}{0.3,0.0,0.8}

\begin{document}

\title{\textbf{Unveiling gravity's quantum fingerprint through gravitational waves}}

\author{$^a$Partha Nandi}
\email{pnandi@sun.ac.za}
\affiliation{$^a$Department of Physics, University of Stellenbosch, Stellenbosch-7600, South Africa}
\author{$^b$Bibhas Ranjan Majhi}
\email{bibhas.majhi@iitg.ac.in}
\affiliation{$^b$Department of Physics, Indian Institute of Technology Guwahati, Guwahati 781039, Assam, India.}

\begin{abstract}
We introduce an innovative method to explore gravity's quantum aspects using a novel theoretical framework. Our model delves into gravity-induced entanglement (GIE) while sidestepping classical communication limitations imposed by the LOCC principle. Specifically, we connect a non-relativistic two-dimensional quantum oscillator detector with linearly polarized gravitational waves (GWs), leveraging the quantum properties inherent in GWs to observe GIE within the oscillator's quantum states. Because our model adheres to both the ``event" and the ``system" localities, the detected GIE serves as a robust indicator of gravity's quantum nature. Detecting this entanglement via gravitational wave detectors could corroborate gravity's quantization and unveil crucial properties of its sources.
  
\end{abstract}

\maketitle
{\it Introduction. --}\label{sec1}
The fusion of gravitation and quantum mechanics into a unified theory, often termed ``quantum gravity" (QG), stands as one of the most tantalizing frontiers in modern physics \cite{PhysRev.162.1239,Penrose:1996cv,Nandi:2023tf}. The absence of experimental evidence testing the quantum nature of gravity makes the undertaking even more daunting.  The lack of a comprehensive theory of QG may underlie our struggles to elucidate enigmatic phenomena such as dark matter, dark energy, and the cosmological constant problem \cite{Kiefer:2004xyv, RevModPhys.61.1}. Despite considerable efforts, a comprehensive theory of quantum gravity remains elusive. String theory \cite{Green:1987sp} has emerged as a notable contender for quantum gravity, yet it grapples with several conceptual obstacles \cite{Smolin:2000af, Rovelli:2014ssa}. However, no experiment or observation has definitively supported or refuted any quantum gravity theory. Given the large energy scale of quantum gravity, referred to as the ``Planck energy scale" at approximately $10^{16}$ TeV, direct tests in colliders are impractical. Therefore, it is essential to search for indirect indications of these theories through attainable, low-energy laboratory experiments \cite{PhysRevLett.101.221301}. Numerous authors have investigated potential quantum gravity signatures in experiments involving condensed matter, atomic, and molecular systems \cite{PhysRevA.102.062807, PhysRevD.84.044013}. However, none have been able to isolate the specific effects of the quantum nature of gravity without resorting to ad hoc exotic assumptions such as the Generalized Uncertainty Principle (GUP) \cite{Pikovski:2011zk}.

In this context, it's intriguing to delve into whether we can propose natural observable consequences of a quantum model that allows us to delve into the quantum aspects of gravity or potentially question them. Our aim in this communication is to present a quantum mechanical approach for discerning the quantum properties of gravitational interaction using a continuous bipartite quantum system.

Quantum information theory establishes that entanglement cannot be generated between two isolated systems through local operations and classical communication (LOCC) \cite{PhysRevA.54.3824}. However, our recent research indicates that classical linearized gravity can induce entanglement in a quantum system \cite{PhysRevLett.119.240401,PhysRevLett.119.240402,Nandi:2024zxp} via a nonlocal interaction of gravitation. This phenomenon, often referred to as Gravitational Induced Entanglement (GIE), suggests that entanglement solely induced by gravity does not conclusively prove the quantum nature of gravitational interaction.
Furthermore, the concept of locality encompasses two crucial perspectives: (a) ``event locality," proposing that operations occur only at spacetime events without affecting causally disconnected events, and (b) ``system locality" \cite{Galley:2020qsf}, grounded in quantum mechanics, suggesting that operations affecting two quantum systems must be separable.

Our objective is to explore whether quantum gravitational interaction, when implemented as a ``local" interaction in a bipartite system Hamiltonian as discussed in \cite{PhysRevD.108.L101702, PhysRevLett.130.100202}, can lead to the emergence of entanglement between subsystems of the overall system Hamiltonian, especially in situations where classical channels for information exchange weren't initially established. This nontrivial entanglement between the two subsystems would be a critical indicator of the quantum nature of the local gravitational interaction.

Here we present a formalism for investigate the unambiguous phenomenon of GIE, circumventing the classical communication (CC) constraints of the LOCC principle. 
The milestone achievement of detecting gravitational waves (GWs) by the LIGO-Virgo collaboration in 2015 \cite{PhysRevLett.116.061102}, along with advancements towards detecting gravitational waves in a stochastic background by LIGO-Virgo-KAGRA \cite{NANOGrav:2023gor}, has paved the way for exploring the quantum mechanics of gravitational interaction \cite{PhysRevLett.127.081602, Parikh:2020fhy,PhysRevD.109.026013}. Therefore GWs provide alternative experimental environment to explore the quantum nature of gravity.
Our proposed model is consists of a non-relativistic two-dimensional quantum harmonic oscillator (QHO), interacting with linearized GWs. Since the detection of gravitational waves occurs at a length scale of the order of $\sim 10^{-18}m$, from the practical point of view, then the quantum nature of the detector model cannot be neglected \cite{RevModPhys.52.341, Thorne:1997ut}. Additionally, various resonant bar detectors have made significant advancements \cite{PhysRevD.47.362, PhysRevLett.74.1908}, demonstrating the capability of modern gravitational wave interferometers to monitor the relative positions of their test masses with an exceptional accuracy of around $10^{-20}m$ \cite{PhysRevD.78.122002}. 

Our model contains important features to fulfil the requirement of system locality. The QHOs are interacting only with ``plus'' polarization of GWs, characterized by a long wavelength and localized at a fixed time $(t)$. For this scenario, the inability of two modes of the system to communicate with each other via local classical gravitational interaction is suggested by the factorization of the time evolution operator into two independent modes (system localized) in the interaction picture: $\hat{U}^{(g)}_{\text{int}}(t) = U^{(g)}_1 (t)\otimes \hat{U}^{(g)}_2(t)$. Within this, considering the GWs as quantum mode, we perturbatively evolve the initial unentangled states of the two HOs. This analysis shows nontrivial entanglement appearing between the two modes of the oscillators. More importantly, the vacuum fluctuation of the quantum modes (gravitons) of GWs is completely responsible for such gravitationally induced entanglement. Therefore if such entanglement between the states of a quantum system through interaction with plus polarised GWs is detectable, then that will lead to a direct establishment of quantum nature of gravity.


{\it Quantum model. --}\label{sec2}
GWs affect the geodesics, and particularly at the linearised level the geodesics is deviated in the perpendicular directions of the propagation of GWs. The whole scenario can be modelled as one dimensional motions of two particles. We may consider these particles are initially trapped in two independent harmonic potentials along these two directions and then GWs are affecting their motions. Then at the weak field limit regime and for the GW is moving along $z$-axis, the whole system can be described by a very simple Hamiltonian (written in terms of operators)
\begin{eqnarray}
\hat{H}&=&\sum_{i=1,2}\Big(\alpha{\hat{P}_i}^2 + \beta{\hat{X}_i}^2\Big)+\gamma(t)\Big(\hat{X}_{1}\hat{P}_{1}+\hat{P}_{1}\hat{X}_{1}\Big) 
\nonumber
\\
&&-\gamma(t)\Big(\hat{X}_{2}\hat{P}_{2}+\hat{P}_{2}\hat{X}_{2}\Big) + \delta(t) \Big(\hat{X}_1\hat{P}_2 + \hat{P}_1 \hat{X}_2\Big)~,
\label{mg}
\end{eqnarray}
where $M$ denotes the particle's mass, $\alpha = \frac{1}{2M}$, $\beta = \frac{1}{2} M\Omega_0^2$, $\gamma(t) = \dot{\chi}(t)\epsilon_+$ and $\delta =2 \dot{\chi}(t)\epsilon_\times$. Here $2\chi(t)$ denotes the time variation of the gravitational wave (GW). In transverse-traceless gauge the GW has been taken as $h_{jk}=2\chi(t)(\epsilon_{\times}\sigma_{1jk}+\epsilon_{+}\sigma_{3jk})$. $\sigma_{1jk}$ is the $(jk)^{th}$ element of the Pauli matrix $\sigma_1$ and so on.

The above model has been proposed long ago \cite{Speliotopoulos:1994xh} and has been in the light of investigation \cite{PhysRevD.108.124069} (more details on the above Hamiltonian and its understanding can be found in \cite{Nandi:2024zxp}). In the above, $\epsilon_{\pm}$ represents plus (cross) polarization. Note that $\epsilon_\times$ provides a coupling among the HOs and since we are interested in whether the local effects on individual HOs can make a quantum communication between them, we will not consider this.  
Note that in equation (\ref{mg}), the classical GWs interaction terms corresponding to the plus polarization are two completely decoupled terms defined at a specific instant ($t$). These are consistent with the general observation that $\epsilon_+$ does not mix the two perpendicular directions while $\epsilon_\times$ does so.  This suggests that the terms corresponding to $\epsilon_+$ are local operators in time, aligning with the concept of event locality. Additionally, in the interaction picture, the time evolution of the system can act separately on a separable Hilbert space associated with the independent modes of the oscillator. In other words, we can factorize the time evolution operator of the composite system as: $\hat{U}^{\gamma}_{\text{int}}(t)=\hat{U}^{\gamma}_{\text{int},1}(t)\otimes \hat{U}^{\gamma}_{\text{int},2}(t)$, thus satisfying system locality. This implies that the time evolution implemented by $U^{\gamma}_{\text{int}}(t)$ cannot create entanglement if the gravitational field lacks quantum degrees of freedom. 
Therefore, for further analysis, we will omit the term involving $\delta$. Here, we will focus solely on the essential quantum effect of the plus polarization by neglecting the cross-polarization terms.  Also note that for the propagation along the $z$-direction, $\gamma(t)$ is a function of $z$; whereas the HOs are along the other two perpendicular directions. Hence, the quantum operators for GWs and those for HOs will be within two independent Hilbert spaces.

To clarify the points mentioned above, let us introduce the annihilation operator for the two-mode harmonic oscillator, defined by
\begin{eqnarray}
&&\hat{a}_{i}=\Big(\frac{\alpha}{\beta}\Big)^{1/4}\Big(\frac{\sqrt{\frac{\beta}{\alpha}}\hat{X}_{i}+i\hat{P}_{i}}{\sqrt{2\hbar}}\Big)~.
\end{eqnarray}
The corresponding creation operator is given by $\hat{{a}}_{i}^{\dagger}$
with $[\hat{a}_{i},\hat{{a}}_{j}^{\dagger}]=\delta_{ij} \mathbb{I};i,j=1,2$. These lead to the following form of the Hamiltonian
\begin{equation}
\hat{H}(t)=\underbrace{2\hbar\sqrt{\alpha\beta}\Big(\sum_{i=1,2} \hat{N}_i+1\Big)}_{H_0}+\underbrace{i\hbar \hat{\gamma}(t)({\hat{a}}{_{1}^{\dagger}}^2 - {\hat{a}}_{1}^2 - {\hat{a}}{_{2}^{\dagger}}^2+\hat{a}_{2}^2)}_{H_{\textrm{int}}(t)}~.
\label{HAMILTONIAN}
\end{equation}
Here we denote $\hat{N}_i = \hat{{a}}_{i}^{\dagger}\hat{a}_{i}$ and we have taken only $\epsilon_+$ polarization into account. Here we  treat $H_{\textrm{int}}(t)$ as perturbation. The above expression shows that the two HOs are interacting with GWs individually and therefore these interactions are local in nature. This does not cause any interaction between the HOs at the classical level. Here we are interested to see whether the quantum nature of $\hat{\gamma}(t)$ can produce any quantum communication between these oscillators.

Since the interaction is time-dependent, the system can be studied within the time-dependent perturbation theory, in which the interaction picture is the suitable formalism. 
Consider that the two-dimensional HO system and the graviton both were initially ($t=0$) in the ground state $\ket{00;0}_{t=0}=(\ket{0}_{1}\otimes\ket{0}_{2})_{HO}\otimes\ket{0}_G$.
Such a choice leads to an initially unentangled state between the HOs.
Then, in the interaction picture, the time evolution of the state is given by 
\begin{equation}
\ket{00;0}_t^I=\hat{U}^{\hat{\gamma}}_{\text{int}}(t,0)\ket{00;0}_{t=0}~,
\label{BRM2}
\end{equation}
where 
${\hat U}^{\hat{\gamma}}_{\text{int}}(t,0)={\hat T}e^{-\frac{i}{\hbar}{\int_{0}^{t}{\hat {H}}_{\textrm{int}}^I(t^\prime)dt^\prime}}$. ${\hat T}$ represents the time-ordered product between the operators. ${\hat {H}}_{\textrm{int}}^I$ is given by 
\begin{eqnarray}
\hat{H}_{\textrm{int}}^I(t)&=&e^{\frac{i\hat{H}_0t}{\hbar}}\hat{H}_{\textrm{int}}(t)e^{-\frac{i\hat{H}_0t}{\hbar}}
\nonumber
\\
&=& \hat{H}_{1,int}-\hat{H}_{2,int},
\nonumber
\\
&=&i\hbar\hat{\gamma}^{I}(t)e^{\frac{i\hat{H}_0t}{\hbar}}({\hat{a}}{_{1}^{\dagger}}^2-\hat{a}_{1}^2-{\hat{a}}{_{2}^{\dagger}}^2+\hat{a}_{2}^2))e^{-\frac{i\hat{H}_0t}{\hbar}}~.
\end{eqnarray}
with $
\hat{H}_{1,\text{int}} = i\hbar \left[ \left(\hat{a}_{1}^{\dagger 2}(t) - \hat{a}_{1}^2(t) \right) \otimes \mathbb{I}_{2} \right] \otimes \hat{\gamma}^{I}(t)$,
and $\hat{H}_{2,\text{int}} = i\hbar \left[ \mathbb{I}_{1} \otimes \left(\hat{a}_{2}^{\dagger 2}(t) - \hat{a}_{2}^2(t) \right) \right] \otimes \hat{\gamma}^{I}(t)$.

The quantum operator $\hat{\gamma}^{I}(t)$ for the GWs in the interaction picture has been obtained in \cite{Parikh:2020fhy}. Following the same procedures as in our case, we find the following:
\begin{equation}
\hat{\gamma}^I(t)=\frac{\dot{\hat{\chi}}}{2}=iC_{\gamma}(\hat{b}e^{-i\omega_gt}-\hat{b}^\dagger e^{i\omega_gt})~.
\label{BRM1}
\end{equation}
Here, $C_{\gamma}=-\sqrt{\frac{\omega_g c \pi l_p^2}{2L^3}}$ represents a constant term, and $\omega_g$ denotes the frequency of the incident gravitational wave, subject to the box quantization with a box length $L$. In this expression, $\hat{b}$ (${\hat{b}}^\dagger$) corresponds to the single-mode annihilation (creation) operator for a graviton. We also present it in the supplemental material to make it consistent with our notations. 
The whole analysis will be done within the second-order perturbation calculation, i.e., up to order $C_\gamma^2$,
and therefore the terms with higher order will be neglected. We will find that retaining up to quadratic terms in the perturbation series leads to a leading-order non-trivial contribution in our desired quantities.

{\it Entanglement phenomenon. --}\label{Sec3}
Now, using the standard techniques and keeping upto second-order terms, one finds
\begin{eqnarray}
&&\ket{00;0}_t^I=(\hat{C}_{00}^{(0)}+\hat{C}_{00}^{(2)})\ket{00;0}+\hat{C}_{02}^{(1)}\ket{02;0}
+\hat{C}_{20}^{(1)}\ket{20;0}
\nonumber
\\
&&+\hat{C}_{04}^{(2)}\ket{04;0}+\hat{C}_{40}^{(2)}\ket{40;0}+\hat{C}_{22}^{(2)}\ket{22;0},
\label{finalstt}
\end{eqnarray}
where $\hat{C}_{00}^{(0)} = 1$ and
\begin{eqnarray}
\hat{C}_{00}^{(2)} &=& -\frac{4iC^2}{\hbar^2}{\int_{0}^{t}}dt_1{\int_{0}^{t_1}}dt_2\hat{\gamma}^I(t_1)\hat{\gamma}^I(t_2)\sin T
\nonumber
\\
&&+\frac{2C^2}{\hbar^2}{\int_{0}^{t}}dt_1{\int_{0}^{t_1}}dt_2[\hat{\gamma}^I(t_1),\hat{\gamma}^I(t_2)]{e^{T_-}}
\nonumber
\\
&&+\frac{2C^2}{\hbar^2}{\int_{0}^{t}}dt_2{\int_{0}^{t}}dt_1\hat{\gamma}^I(t_2)\hat{\gamma}^I(t_1){e^{T_-}}~;
\nonumber
\\
\hat{C}_{02}^{(1)}&=&\frac{\sqrt{2}iC}{\hbar}{\int_{0}^{t}}dt_1\hat{\gamma}^I(t_1){e^{2i\Omega_0 t_1}}~;
\nonumber
\\
\hat{C}_{20}^{(1)}&=&-\frac{\sqrt{2}iC}{\hbar}{\int_{0}^{t}}dt_1\hat{\gamma}^I(t_1){e^{2i\Omega_0 t_1}}~; 
\nonumber
\\
\hat{C}_{04}^{(2)} &=&-\frac{\sqrt{6}C^2}{\hbar^2}{\int_{0}^{t}}dt_1{\int_{0}^{t_1}}dt_2[\hat{\gamma}^I(t_1),\hat{\gamma}^I(t_2)]{e^{T_+}}
\nonumber
\\
&&-\frac{\sqrt{6}C^2}{\hbar^2}{\int_{0}^{t}}dt_2{\int_{0}^{t}}dt_1\hat{\gamma}^I(t_2)\hat{\gamma}^I(t_1){e^{T_+}}~;
\nonumber
\\
\hat{C}_{40}^{(2)}&=&-\frac{\sqrt{6}C^2}{\hbar^2}{\int_{0}^{t}}dt_1{\int_{0}^{t_1}}dt_2[\hat{\gamma}^I(t_1),\hat{\gamma}^I(t_2)]{e^{T_+}}
\nonumber
\\
&&-\frac{\sqrt{6}C^2}{\hbar^2}{\int_{0}^{t}}dt_2{\int_{0}^{t}}dt_1\hat{\gamma}^I(t_2)\hat{\gamma}^I(t_1){e^{T_+}}~; 
\nonumber
\\
{\hat{C}}_{22}^{(2)} &=& \frac{2C^2}{\hbar^2}{\int_{0}^{t}}dt_1{\int_{0}^{t_1}}dt_2[\hat{\gamma}^I(t_1),\hat{\gamma}^I(t_2)]{e^{T_+}}
\nonumber
\\
&+&\frac{2C^2}{\hbar^2}{\int_{0}^{t}}dt_2{\int_{0}^{t}}dt_1\hat{\gamma}^I(t_2)\hat{\gamma}^I(t_1){e^{T_+}}~.
\end{eqnarray}
Here we denote $C=i\hbar$, $T=2\Omega_0(t_1-t_2)$, $T_+=2i\Omega_0(t_1+t_2)$ and $T_-=2i\Omega_0(t_1-t_2)=iT$. For completeness, the steps to obtain (\ref{finalstt}) are given in the Supplementary Material.

Before proceeding, note that the state (\ref{finalstt}) can also be expressed in the following fashion:
\begin{eqnarray}
&&\ket{00;0}_{t}^I = \left[(\mathbb{I}_{1} \otimes \mathbb{I}_{2} \otimes \sqrt{\hat{B}_{1}})|0\rangle_{1} - (\mathbb{I}_{1} \otimes \mathbb{I}_{2} \otimes \sqrt{\hat{B}_{2}})|2\rangle_{1} \right. \nonumber \\
&& \left. + (\mathbb{I}_{1} \otimes \mathbb{I}_{2} \otimes \sqrt{\hat{B}_{3}})|4\rangle_{1}\right] \otimes \left[(\mathbb{I}_{2} \otimes \sqrt{\hat{B}_{1}})|0\rangle_{2} \right. \nonumber \\
&& \left. + (\mathbb{I}_{2} \otimes \sqrt{\hat{B}_{2}})|2\rangle_{2} + (\mathbb{I}_{2} \otimes \sqrt{\hat{B}_{3}})|4\rangle_{2}\right] \otimes |0\rangle_{G}~,
\label{PLB2}
\end{eqnarray}
\normalsize
with \( \hat{B}_1 = (\hat{C}_{00}^{(0)}+\hat{C}_{00}^{(2)}) \), \( \hat{B}_2 = - \hat{C}_{22}^{(2)} \), \( \sqrt{\hat{B}_1\hat{B}_2} = \hat{C}_{02}^{(1)} = - \hat{C}_{20}^{(1)} \), and \( \sqrt{\hat{B}_1\hat{B}_3} = \hat{C}_{40}^{(2)} = \hat{C}_{04}^{(2)} \). Using these definitions, it can be seen that \( \sqrt{\hat{B}_3} \) is of the second order in perturbation. Consequently, (\ref{PLB2}) simplifies to (\ref{finalstt}) up to the second order in perturbation. This form highlights the non-separable structure between the oscillatory subsystems 1, 2, and the graviton environment (ancillary system) which is only due to the non-trivial action of the $\sqrt{\hat{B}_{i}}$ with $i=1,2,3$ dependent terms on $\ket{0}_{G}$.

At this stage, it is worthwhile to mention that even if the time evolution operator ${\hat U}^{\hat{\gamma}}_{\text{int}}(t,0)$ of the composite system is factorized, this will still generate non-separable final states through unitary evolution in systems with initially separable states ($(\ket{0}_{1}\otimes\ket{0}_{2})\otimes\ket{0}_{G})$. This may appear surprising at first glance, but a careful analysis reveals that the factorized form of the time evolution $\hat{U}^{\gamma}_{\text{int}}(t)=\hat{U}^{\gamma}_{\text{int},1}(t)\otimes \hat{U}^{\gamma}_{\text{int},2}(t)$ indicates that even though $\hat{U}^{\gamma}_{\text{int},1}$ and $\hat{U}^{\gamma}_{\text{int},2}(t)$ act independently on $\ket{0}_{1}$ and $\ket{0}_{2}$, they both involve $\hat{\gamma}^I(t)$ through $\hat{H}_{1,\text{int}}$ and $\hat{H}_{2,\text{int}}$. This shared dependency on the common quantum (graviton) environment, manifested through $\hat{\gamma}^I(t)$, generates correlations between the two oscillatory modes $\hat{a}_{1}$ and $\hat{a}_{2}$ during the unitary evolution, leading to the nonseparable nature of the final state. The key principle here is that factorization of the unitary evolution operator alone does not guarantee that the final state will remain disentangled (separable slot by slot by tensor product associated with individual Hilbert spaces) if the systems share interactions. Actually, the fundamental idea is that nonseparable behavior of the final state can be generated through interactions between subsystems via a common quantum environment, even when there are no direct interactions between the subsystems (two oscillator modes). This remains true even if the initial state is separable and the evolution operator can be factorized, as discussed in detail in \cite{RevModPhys.75.715,PhysRevLett.91.037902,PhysRevLett.89.277901,PhysRevA.73.062306}.

Moreover, it should be noted that under unitary time evolution, the density matrix corresponding to the final state remains pure as desired. However, this does not imply that the final state is factorizable, as is the case with the initial state of the composite system. The nonseparable nature of the final state prompts us to mention that the graviton environment can induce a nontrivial entanglement dynamics between two independent modes of the oscillator-detector system in the absence of any direct interaction between the two modes. However the quantification of the entanglement can be done as follows. The density matrix is given by $
\hat{\rho}_f(t) = \ket{00;0}_t^I \bra{00;0}_t^I$.
Since our focus is solely on the harmonic oscillators, with gravitons serving as background objects, we trace out the graviton sector from the density matrix.  This yields $\hat{\rho}_{12}(t)=\textrm{Tr}_G\hat{\rho}_f(t)=\sum_{n}{_G}\bra{n}\hat{\rho}_f(t)\ket{n}_{G}$.
Finally, the reduced density matrix of any one mode of the oscillator (say, the first mode) is then obtained as
\begin{eqnarray}
\hat{\rho}_1(t) &=& \textrm{Tr}_2\hat{\rho}_{12}(t)=\sum_{n_2=0}^\infty\bra{n_2}\hat{\rho}_{12}(t)\ket{n_2}
\nonumber
\\
&&=K_{00}\ket{0}\bra{0}+K_{22}\ket{2}\bra{2}+K_{40}\ket{4}\bra{0}
\nonumber
\\
&&+K_{40}^*\ket{0}\bra{4}~,
\label{redden}
\end{eqnarray}
with $K_{00}=\big<1+\hat{C}_{00}^{(2)}+\hat{C}{_{00}^{(2)}}^{\dagger}+\hat{C}{_{02}^{(1)}}^{\dagger}\hat{C}_{02}^{(1)}\big>_G$; $K_{22}=\big<\hat{C}{_{20}^{(1)}}^{\dagger}\hat{C}_{20}^{(1)}\big>_G$; $K_{40}=\big<\hat{C}_{40}^{(2)}\big>_G$; $K_{40}^*=\big<\hat{C}{_{40}^{(2)}}^{\dagger}\big>_G$.
Here, the symbol $\big<\hat{A}\big>_G$ denotes the vacuum expectation value of the operator $\hat{A}$
with respect to the graviton sector, i.e, ${_G}\bra{0}\hat{A}\ket{0}_G$.
In the above, $\textrm{Tr}_2$ represents the tracing over the quantum states of the second mode of the oscillator.
In order to investigate whether the final state (\ref{finalstt}) represents an entangled state, we will study two quantities that can quantify the entanglement between the quantum systems. These are von Neumann entropy and purity.

The later time entropy of the first mode of the oscillator can be determined by the formula $S(t) = -\textrm{Tr}_1\Big(\hat{\rho_1}(t)\ln\hat{\rho}_1(t)\Big)$. From (\ref{redden}), this is evaluated as
\begin{equation}
S(t)=-K_{00}\ln K_{00}-K_{22}\ln K_{22}~.
\end{equation}
Using the explicit structures of $K_{00}$ and $K_{22}$, one obtains $K_{00}=1-K^{(2)}$; $K_{22}=K^{(2)}$ with 
\begin{equation}
K^{(2)}=2{\int_{0}^{t}}dt_1{\int_{0}^{t}}dt_2{e^{T_-}}\big<\hat{\gamma}^I(t_2)\hat{\gamma}^I(t_1)\big>~.
\label{vac}
\end{equation} 
Then using (\ref{BRM1}) and keeping terms only upto second order, the von-Neumann entropy takes the form
$S(t) \simeq K^{(2)}(t)$
where
\begin{equation} 
K^{(2)}(t) = \frac{8C_{\gamma}^2}{\Omega^2}\sin^2\Big(\frac{\Omega t}{2}\Big)~.
\label{B1}
\end{equation}
with $\Omega=2\Omega_0+\omega_g$.
Note that $K^{(2)}(t)$ is always positive.
It is evident that at the initial time $t=0$, $S(t)$ was zero, but it can be non-vanishing at a later time. Thus it indicates that the HOs are now entangled.

The Purity function is defined as $P(t) = \textrm{Tr}_1(\hat{\rho}_1^2(t))$. In our case, this can be calculated as
\begin{eqnarray}
P(t)\simeq K_{00}^2
\simeq 1-2K^{(2)}(t)~.
\label{lk}
\end{eqnarray}
So the Purity function is departing from unity and this is the footprint of creating entanglement between the two modes of the HO.

The important observation is that the entanglement is due to the vacuum fluctuation of the gravitational modes (see Eq. (\ref{vac})). To quantify it, concentrated on $\big<\hat{\gamma}^I(t_2)\hat{\gamma}^I(t_1)\big>$. Note that $\hat{\gamma}^I(t_2)\hat{\gamma}^I(t_1)$ can be divided into two parts: $(1/2) [\hat{\gamma}^I(t_2), \hat{\gamma}^I(t_1)] + (1/2)\{\hat{\gamma}^I(t_2), \hat{\gamma}^I(t_1)\}$. Among these two, the commutator part provides vacuum independent value $2iC_{\gamma}^2 \sin[\omega_g(t_1-t_2)]$, while the anti-commutator part depends on the state for which the expectation value is evaluated. Therefore the anti-commutator is the actual quantity which quantifies the role of vacuum fluctuation of GWs modes to the entanglement. This for the present case in vacuum gives $2C_\gamma^2 \cos[\omega_g(t_1-t_2)]$.

It's important to note at this stage that equation (\ref{lk}) represents a periodic function of time, which underpins the dynamic nature of the degree of entanglement and gives rise to the phenomena known as entanglement revival and collapse \cite{PhysRevLett.65.3385}. When there is entanglement between the quantum states of two systems, any change in one system induces a quantum change in the other. In the present case, the conditions for maximum and no-entanglement (for a non-zero finite value of interaction time $t$) are $\Omega t = (2n+1)\pi$ and $\Omega t = (n+1)2\pi$, respectively where $n=0,1,2,\dots$ (see Eq. (\ref{B1})). Particularly intriguing is the no-entanglement condition. Once entanglement is established between the two harmonic oscillators (HOs), adjusting the natural frequency $\Omega_0$
  of the HOs allows one to achieve the no-entanglement condition, ensuring that changes in one HO do not affect the quantum state of the other. This not only confirms the observation of gravitational induced entanglement (GIE) but also determines the frequency of the gravitational waves (GWs).

Furthermore, we have noticed that the quantum effects of the environment play a critical role in our formulation. The time evolution operator can be factorized into two separate evolution operators corresponding to the individual oscillatory modes of the detector. Each mode is associated with an operator-valued quantum field linked to background quantum gravitational waves (GWs), which act as a quantum environment. However, the mere factorization of the unitary evolution operator does not ensure that the final state will remain separable if the subsystems interact through a shared quantum environment (see, for instance, \cite{Breuer:2007juk}). For a classical background gravitational field, the final states remain separable since the operator status of $\sqrt{\hat{B}_i}$ is inactive and behaves as classical parameters. In this scenario, the concept of $\ket{0}_{G}$ is irrelevant, and there is no direct interaction between the two modes of the oscillator. Hence, a classical gravitational environment does not exhibit any entanglement. However, for a quantum environment, the nonseparability of the final state, evidenced by the loss of purity in the subsystem, is a natural consequence of these interactions. Although the time evolution is unitary and the final state of the composite system must be pure (see the discussion around Eq. (\ref{PLB2})), our focus is not on the composite system but on the properties of the HOs' modes, initially separable. According to open quantum system formalism \cite{Breuer:2007juk}, the degrees of freedom of the environment must be traced out. The initial state $\ket{0,0;0}_{t=0} = \ket{0,0}\otimes \ket{0}_{G}$ shows that the HO modes are separable even after tracing out the $G$-modes. However, in the final state, the HOs are not separable once the $G$-modes are traced out. If we calculate the reduced density matrix after tracing out the unobserved graviton modes (as our detector cannot directly detect gravitons, only experiencing indirect effects) and one of the detector's modes, we obtain a mixed state. The emergence of nontrivial entanglement entropy indicates that the quantum modes of GWs can produce entanglement between two classically decoupled modes of the oscillatory detector. The environment affects our system, and once the unobserved quantum degrees of freedom are traced out, their impact is visible on our system. We are examining a part of the composite system, and the unobserved information is reflected as entanglement entropy in the system under investigation. This scenario is analogous to the physics of Unruh-DeWitt (UD) detectors (atoms) and their potential to probe the Unruh effect. There is current interest in determining whether the quantum fluctuation of fields (regarded as an environment) can produce entanglement between two causally disconnected UD detectors (see e.g. \cite{Reznik:2002fz, Reznik:2003mnx, Martin-Martinez:2015qwa, Koga:2018the, Koga:2019fqh, Pozas-Kerstjens:2015gta, Barman:2021bbw, Barman:2021kwg, Chowdhury:2021ieg, Barman:2022xht, Barman:2023rhd, Barman:2023wkr} and references therein). The same formalism and ideas are adopted in these investigations as well.

{\it Discussion. --}\label{Sec4}
To summarize the significant findings of the paper and compare them with existing literature, we focus on GIE based on LOCC principle, particularly concerning the entanglement channel between two massive objects due to gravity. Our study demonstrates the entanglement between two independent modes of the oscillator detector, where we model the suspended mirror of a gravitational wave interferometer as an oscillating point particle relative to the long wavelength of the quantum modes of gravitational waves. This GIE represents a key hallmark of the quantum nature of linearized gravity and holds potential for measurement at gravitational wave detectors.

Previous references  \cite{PhysRevLett.119.240401, PhysRevLett.119.240402} lack a transparent argument for studying gravity-mediated entanglement. BMV-like models proposed by Bose, Marletto, and Vedral, where two massive objects interact through classical Newtonian gravitational coupling, are relativistically local. However, when treated as quantum operators involving the coordinates of the massive objects, they become nonlocal operators. Therefore, claiming the quantum nature of gravity based on LOCC arguments is not straightforward \cite{PhysRevD.108.L101702}.

In our paper, we introduce a formulation to investigate GIE resulting from the interaction of a HO system with linearly polarized GWs. Our central points are twofold:

(i) In the context of classical GWs and a quantum HO model, the interaction term remains locally observable in both event locality and system locality. Thus, according to LOCC principles, it cannot generate entanglement between the two modes of the quantum oscillator detector model.

(ii) However, upon quantizing GWs and coupling each mode of the oscillator with single modes of quantized GWs, we observe gravity-induced entanglement at the second order of gravitational perturbation. Notably, this entanglement arises purely from vacuum fluctuations of the quantized GW field, as it depends on the two-point correlation function of quantized GWs.

It may be mentioned that GWs exhibit both plus and cross polarizations when interacting with nonrelativistic masses. These two polarization modes can be investigated separately due to their orthogonality. 
In this regard remember that even in text book the effects of these polarization on a classical system have been investigated separately to know their individual role (e.g. see \cite{SMC}). This is the usual practice.
Recently, Parikh et al. \cite{PhysRevLett.127.081602, Parikh:2020fhy} also concentrated on the plus polarization modes of gravitational waves  only to explore their quantum mechanics. The primary goal of our paper is to illustrate the quantum nature of gravitational waves through nontrivial entanglement dynamics that lack classical counterparts.
However, in our recent work \cite{PhysRevD.108.124069}, we demonstrated that by rotating the axis of linearly cross-polarized gravitational waves to align with the axis of the plus polarization, we can effectively reduce the system of 2D harmonic oscillator modes. This alignment allows each mode to interact individually with an effective plus polarization mode along the rotated new axis.
Following this sprit and aiming to know the role of individual polarizations on the gravitationally induced entanglement, we prefer to investigate them separately. \cite{Nandi:2024zxp} previously investigated the effect of cross-polarization on a quantum system. Therefore, we concentrate here on the plus polarization.  Looking at the notable difference from the other polarization (plus polarization, unlike cross one, does not produce classical communication between the HOs), it is very important to investigate its quantum nature. We particularly did this in the present study and found notable properties that are distinct from those of the cross-poralization part. We see that although plus polarization at the classical level does not produce any communication, it is capable of producing the same at the quantum scale. However the cross polarization does the same at both the regime.

Our results indicate that local classical communication due to gravity (GWs) cannot create entanglement, but local quantum communication can. This distinction underscores the intricate interplay between the classical and quantum aspects of gravitational interactions and their implications for understanding the quantum nature of gravity.

\section*{Acknowledgements}
One of the authors (PN) acknowledges the support of
postdoctoral fellowship grants from Stellenbosch University, South Africa.  We would also like to extend our thanks to Nandita Debnath for carefully reading the manuscript and for providing valuable suggestions and comments during the course of this work. The authors thank the referees for valuable comments and
suggestions.

\bibliographystyle{apsrev}
\bibliography{bibtexfile1}


\clearpage

\pagebreak
\widetext
\begin{center}
\textbf{\huge{Supplemental Materials}\\
\vskip 5mm
 \large Unveiling gravity's quantum fingerprint through gravitational waves}\\
Partha Nandi$^a$ and Bibhas Ranjan Majhi$^b$\\
{\it $^a$Department of Physics, University of Stellenbosch, Stellenbosch-7600, South Africa.\\
$^b$Department of Physics, Indian Institute of Technology Guwahati, Guwahati 781039, Assam, India.}
\end{center}

\email{parthanandyphysics@gmail.com}
\email{bibhas.majhi@iitg.ac.in}


\setcounter{equation}{0}
\setcounter{figure}{0}
\setcounter{table}{0}
\setcounter{page}{1}
\makeatletter
\renewcommand{\theequation}{A\arabic{equation}}
\renewcommand{\thefigure}{S\arabic{figure}}
\renewcommand{\bibnumfmt}[1]{[S#1]}
\renewcommand{\citenumfont}[1]{#1}

\section*{Section I: Quantum GWs}
The analysis is being followed from \cite{PhysRevLett.127.081602,Parikh:2020fhy}.
The linearized version of Einstein-Hilbert action is
\begin{equation}
S_{EH}=-\frac{c^3}{64\pi G}\int d^4x 
 \partial_{\mu}h_{ij} \partial^{\mu}h^{ij}~,
 \label{SEH}
\end{equation}
where $c$ is the speed of light in vacuum and $G$ represents the standard Newtonian gravitational constant. From dimensional analysis, it is clear that our $h_{ij}$ is dimensionless. Under the TT (Transverse Traceless) gauge, $h_{ij}$ can be identified as
\begin{equation}
h_{ij}=\frac{1}{l_p}\sum_{\Vec{k},s}q_{\Vec{k},s}(t)e^{i\Vec{k}.\Vec{x}}\epsilon_{ij}^s(\Vec{k})~.
\end{equation}
Here, $l_p$ denotes the Planck length scale given by $\sqrt{\frac{\hbar G}{c^3}}$ and $s$ represents the polarization of the incident GWs.
Thus, the Einstein-Hilbert action boils down to
\begin{eqnarray}
S_{EH}&=&\frac{c^2L^3}{32\pi Gl_p^2}\int dt \sum_{\Vec{k},s}(\mid \Dot{q}_{\Vec{k},s}\mid^2-\omega_{g}^2\mid {q}_{\Vec{k},s}\mid^2)
\nonumber
\\
&&=\frac{m}{2}\int dt \sum_{\Vec{k}}(\mid \Dot{q}_{\Vec{k}}\mid^2-\omega_{g}^2\mid {q}_{\Vec{k}}\mid^2)~,
\end{eqnarray}
where $m=\frac{c^2L^3}{16\pi Gl_p^2}$ is the mass, $\omega_g = |\vec{k}|$ and $L$ is the artificial length scale i.e. we are making the GWs quantized
assuming that the Gravitons are bounded within a one-dimensional box of length $L$ which imposes the energy quantization on them. Only plus polarization is considered here i.e. $s=+$ and we have used the sum rule of polarization tensors
\begin{equation}
\epsilon_{ij}^s\epsilon^{ij}_{s^\prime}=2\delta^s_{s^\prime}~.
\end{equation}
Therefore, it is evident that each mode of the GWs can be treated as a harmonic oscillator (HO) with mass $m$ and frequency $\omega_g$.

Now for the chosen GWs and for each mode (i.e.  for each $\vec{k}$) we have 
\begin{equation}
h_{ij}(t) = 2\chi(t) \epsilon_{+}\sigma_{3jk}=\frac{1}{l_p} q_{\Vec{k}}(t)e^{i\Vec{k}.\Vec{x}}\epsilon_{ij}(\Vec{k})~.
\end{equation} 
Then we get
\begin{equation}
\chi(t)=\frac{1}{2l_p}q(t)e^{i\Vec{k}.\Vec{x}}~,
\end{equation}
which becomes $\chi(t)=\frac{1}{2l_p}q(t)e^{ikz}$ for GWs moving along the $z$-direction. Here we removed the suffix $\vec{k}$ of $q(t)$. As each $q(t)$ represents a HO with mass $m$ and frequency $\omega_g$, we can write $q(t) = \sqrt{\frac{\hbar}{2m\omega_g}}(\hat{b}e^{-i\omega_gt}+\hat{b}^\dagger e^{i\omega_gt})$.
Thereby, elevating $\chi(t)$ to the quantum status and focusing on its real part for $z=0$, we obtain
\begin{equation}
\hat{\chi}(t)=\frac{1}{2l_p}\sqrt{\frac{\hbar}{2m\omega_g}}(\hat{b}e^{-i\omega_gt}+\hat{b}^\dagger e^{i\omega_gt})~.
\end{equation}
This yields Eq. (\ref{BRM1}).

\section*{Section II: Septs to obtain Eq. (\ref{finalstt})}
Using the Magnus expansion formula \cite{Magnus:1954zz}, ${\hat U^{\hat{\gamma}}_{\text{int}}(t,0)}$ can be computed as 
\begin{eqnarray}
&&\hat U^{\hat{\gamma}}_{\text{int}}(t,0) \simeq\exp\Big(-\frac{i}{\hbar}{\int_{0}^{t}{\hat {H}}_{\textrm{int}}^I(t_1)dt_1}
-\frac{1}{2!\hbar^2}{\int_{0}^{t}dt_1{\int_{0}^{t_1}dt_2[{\hat {H}}_{\textrm{int}}^I(t_1),{\hat {H}}_{\textrm{int}}^I(t_2)]}} \Big)
\nonumber
\\
&=& 1-\frac{i}{\hbar}{\int_{0}^{t}{\hat {H}}_{\textrm{int}}^I(t_1)dt_1} -\frac{1}{2!\hbar^2}{\int_{0}^{t}dt_1{\int_{0}^{t_1}dt_2[{\hat {H}}_{\textrm{int}}^I(t_1),{\hat {H}}_{\textrm{int}}^I(t_2)]}}
-\frac{1}{2\hbar^2}({\int_{0}^{t}{\hat {H}}_{\textrm{int}}^I(t_1)dt_1})^2~.
\label{teo}
\end{eqnarray}
The commutator of the above can be evaluated as
\begin{eqnarray}
[{\hat {H}}_{\textrm{int}}^I(t_1),{\hat {H}}_{\textrm{int}}^I(t_2)]&=&8iC^2\hat{\gamma}^I(t_1)\hat{\gamma}^I(t_2)(\hat{N}_1+\hat{N}_2+1)\sin T
+C^2[\hat{\gamma}^I(t_1),\hat{\gamma}^I(t_2)]
\nonumber
\\
&&\times [{e^{T_+}}({\hat{a}}{_{1}^{\dagger}}^4+{\hat{a}}{_{2}^{\dagger}}^4
-2{\hat{a}}{_{1}^{\dagger}}^2 {\hat{a}}{_{2}^{\dagger}}^2)
-{e^{T_-}}(\hat{a}_{1}^2 {\hat{a}}{_{1}^{\dagger}}^2+\hat{a}_{2}^2 {\hat{a}}{_{2}^{\dagger}}^2)]~,
\label{com}
\end{eqnarray}
with $C=i\hbar$, $T=2\Omega_0(t_1-t_2)$, $T_+=2i\Omega_0(t_1+t_2)$ and $T_-=2i\Omega_0(t_1-t_2)=iT$. We have also used the time-evolved annihilation and creation operators as $\hat{a}(t)=\hat{a}e^{-i\Omega_0 t}$ and $\hat{a}^\dagger(t)=\hat{a}^\dagger e^{+i\Omega_0 t}$. In the above, terms which will not contribute upon acting on the states, have been ignored. Using this in (\ref{teo}) and then applying in (\ref{BRM2}) one obtains (\ref{finalstt}).



\end{document}